\documentclass[]{aastex701}
\usepackage{multirow}% 
\usepackage{booktabs}%

\newcommand{\secdot}{\rlap{.}^{\rm s}}
\newcommand{\arcsecdot}{\rlap{.}^{''}}
\begin{document}

\title{A large misalignment between continuous jet and discrete ejecta in microquasar GRS\,1915+105 during its obscured phase}

%% To identify a corresponding author, use the \correspondingauthor command.
%% The command appends "Corresponding Author: " to the argument it appears at
%% the bottom left of the first page like the output from \email. 

\author[0000-0001-7369-3539]{Wu Jiang}
\affiliation{Shanghai Astronomical Observatory, Chinese Academy of Sciences, 80 Nandan Road, Shanghai 200030, People's Republic of China}
%\affiliation{State Key Laboratory of Radio Astronomy and Technology, A20 Datun Road, Chaoyang District, Beijing, 100101, People's Republic of China}
\email[show]{jiangwu@shao.ac.cn}
\correspondingauthor{Wu Jiang, Zhen Yan}

\author[0009-0003-6680-1628]{Xi Yan}
\affiliation{Xinjiang Astronomical Observatory, Chinese Academy of Sciences, 150 Science-1 Street, Urumqi 830011, People's Republic of China}
\email{}

\author[0000-0002-5385-9586]{Zhen Yan}
\affiliation{Shanghai Astronomical Observatory, Chinese Academy of Sciences, 80 Nandan Road, Shanghai 200030, People's Republic of China}
\email[show]{zyan@shao.ac.cn}

\author[0000-0002-7329-9344]{Ya-Ping Li}
\affiliation{Shanghai Astronomical Observatory, Chinese Academy of Sciences, 80 Nandan Road, Shanghai 200030, People's Republic of China}
%\affiliation{State Key Laboratory of Radio Astronomy and Technology, A20 Datun Road, Chaoyang District, Beijing, 100101, People's Republic of China}
\email{}

\author[0000-0003-0721-5509]{Lang Cui}
\affiliation{Xinjiang Astronomical Observatory, Chinese Academy of Sciences, 150 Science-1 Street, Urumqi 830011, People's Republic of China}
\affiliation{Xinjiang Key Laboratory of Radio Astrophysics, 150 Science 1-Street, Urumqi 830011, People's Republic of China}
\email{}

\author[0000-0003-3540-8746]{Zhi-Qiang Shen}
\affiliation{Shanghai Astronomical Observatory, Chinese Academy of Sciences, 80 Nandan Road, Shanghai 200030, People's Republic of China}
%\affiliation{State Key Laboratory of Radio Astronomy and Technology, A20 Datun Road, Chaoyang District, Beijing, 100101, People's Republic of China}
\email{}

\begin{abstract}
We report a large misalignment between the continuous jet and the discrete ejecta in GRS\,1915+105, detected in April 2023 with the East Asian VLBI network (EAVN). Two-sided ejecta are shown at 6.7\,GHz images and central continuous jets are resolved at 43\,GHz by EAVN quasi-simultaneously. While the continuous jet was aligned with the long‑standing jet position angle (PA) of $\sim147^{\circ}$, the discrete ejecta appeared at a markedly different PA $\sim188^{\circ}$, with the lowest intrinsic velocity $\sim0.35$\,c ever reported. A similar misalignment of PA between discrete ejecta and continuous jet was independently detected in late September of 2023 during a consecutive flare event.
The pronounced and recurrent angular deviations suggest a time-variable jet launching geometry, which, in conjunction with the observed X-ray obscuration, can be attributable to a warped accretion disk. Our result could offer new insight into the fundamental differences between continuous jet and discrete ejecta, and broadly provides a clue to understand the phenomena for transient black hole X-Ray binaries and changing-look active galactic nuclei during the X-ray obscured phase. 

\end{abstract}



\section{Introduction} 

The Galactic low-mass X-ray binary (LMXB) GRS\,1915+105 is a well-known microquasar, notable for its superluminal jets since its discovery in 1992 \citep{Castro-Tirado_1992IAUC.5590....2C, Mirabel_1994Natur.371...46M}. It is located at a distance of 9.4$\pm$0.8 kpc \citep{Reid_2023ApJ...959...85R}, with 1\,mas corresponding to a projected physical size of 9.4 astronomical units (AU). In addition, it has been reported that this system contains a rapidly spinning black-hole (BH) with a mass of 11.2$^{+2.0}_{-1.8}$\,$M_{\odot}$ \citep{Reid_2023ApJ...959...85R,Reid_2014}, accompanied by an evolved donor star with a mass of 0.47$\pm$0.27\,$M_{\odot}$ \citep{Steeghs2013}. 

Among known BH-LMXBs, GRS\,1915+105 has the longest orbital period (33.85$\pm$0.16 days), leading to the formation of an unusually large and luminous accretion disk \citep{Truss2006,ZZ2017MNRAS}. This disk is thought to fuel an extraordinary outburst that has persisted for over three decades since 1992. During this prolonged activity, the source has exhibited strong and highly variable radio emission originating from its relativistic jet \citep[see reviews in ][]{Fender_2004ARA&A..42..317F}. As a result, GRS\,1915+105 has been the subject of extensive radio observations at various spatial resolutions, making its jet one of the best-studied among stellar-mass BH systems. 

Radio imaging of the microquasar GRS\,1915+105 with very long baseline interferometry (VLBI) resolves the nucleus as a compact jet of length at AU scale,  which is the flat-spectrum radio core of accreting black hole and defined as the compact continuous jet \citep{Dhawan_2000ApJ...543..373D}. During a flare event, discrete ejecta are expelled to thousands of AU and fade over several days. The discrete ejecta are optically thin and are often called transient jet in the literature \citep{Dhawan_2000ApJ...543..373D}.

Long-term radio interferometric observations over the past three decades have provided consistent measurements of the jet’s trajectory. Two key angular parameters describing the jet trajectory are derived from these observations: the position angle (PA), defined as the angle of jet axis measured east of north projected on the plane of the sky, and the viewing angle (VA), defined as the angle between the jet axis and the observer’s line of sight. Both the continuous jet and discrete ejecta exhibit a stable PA of $147^\circ\pm8^\circ$ \citep{Rodr_2025ApJ...986..108R}. While the mean VA measured from the discrete ejecta is estimated to be $64^\circ\pm4^\circ$ \citep{{Mirabel_1994Natur.371...46M,Fender_1999MNRAS.304..865F,Reid_2014,Reid_2023ApJ...959...85R}}. On much longer timescales ($\sim 10^{5}$ yr), the jet position angle is inferred to be $\sim 157^\circ$ from the orientation of large-scale jet-interstellar medium interaction sites \citep{Chaty_2001A&A...366.1035C,Tetarenko2018MNRAS,Motta2025A&A}, and is broadly consistent with the values measured over the past decades. 

Despite decades-long studies, connections between the physics of black
holes and the processes underpinning the formation and launch of these
jets remain elusive. One main reason is that continuous jet and transient jet come from different states of BH binary system and thus are rarely concomitant \citep{Fender2004MNRAS,Remillard_2006ARA&A..44...49R}. In early April 2023, a bright radio flare was reported in the BH-LMXB GRS\,1915+105 \citep{Trushkin_2023ATel15974....1T}. Our timely follow-up observations with the East Asian VLBI Network (EAVN) clearly resolved both a compact central continuous jet and two-sided discrete ejecta. Furthermore, the two type jets presented in dramatically different position angles. It provides an opportunity to decipher the mechanism of formation and launching of the two kinds of jets. Detailed information on observations and data reduction are in Section 2. The results and discussions are presented in Section 3 and Section 4, followed by a short summary in the last.

\section{Observations and Data Reduction} \label{section 2}
GRS\,1915+105 exhibited a strong radio flare in April 2023, which peaked around Modified Julian Date (MJD) $\sim$ 60038 with a flux density of $\sim 2.0$\,Jy at 4.7\,GHz \citep{Trushkin_2023ATel15974....1T}. Following this flare, we promptly conducted VLBI observations with the EAVN at 6.7 and 43\,GHz. 

Approximately four months later, a second extremely bright flare was observed in August by \citep{Trushkin_2023ATel16168....1T}, %\footnote{See also \url{https://www.sao.ru/Doc-en/SciNews/2023/Trushkin}.}, 
peaking around MJD~$\sim$~60161 with a flux density of $\sim 5.5$\,Jy at 2.3 GHz. We also analyzed the Very Large Array (VLA) data after this flare in September, retrieved from the NRAO Data Archive (\url{https://data.nrao.edu/}).

\begin{table*}
\footnotesize
\caption{Summary of GRS\,1915+105 Observations} \label{table:GRS1915_observation_summary}
\begin{tabular*}{\textwidth}{@{\extracolsep\fill}ccccccccccccccc}
\hline
Epoch & Stations & $\nu$  & B.W. & $T_{\rm int}$ & Synthesized Beam & $I_{\rm peak}$ & $I_{\rm rms}$ & $S_{\rm tot}$\\
&  &(GHz) & (MHz) & (min) & (mas $\times$ mas, deg) & \multicolumn{2}{c}{(mJy\,beam$^{-1}$)} & (mJy) \\
(1) & (2) & (3) & (4) & (5) & (6) & (7) & (8) & (9) \\
%2019/10/12  & VLBA  & 8.4  & 512 & $\sim260$ & $ 2.17  \times 0.96, -5.3$ &  2.0 & 0.03 & $5.8 \pm 0.6$\\
\hline
2023/04/15 (60049.8)& EAVN  & 6.7 & 256 &  $\sim220$  & $4.89\times1.62$, 17.2 & 165.0 & 0.22  & $484\pm48$ \\
2023/04/17 (60051.8) & EAVN  & 6.7 & 256 & $\sim220$ & $6.42\times2.60, -17.1$ & 194.9 & 0.20 & $414\pm41$ \\
2023/04/23 (60057.8) & EAVN & 6.7 & 256 & $\sim220$ & $ 6.48 \times 2.67, -24.7$ & 145.0 &  0.28 & $296\pm30$ \\
2023/04/16 (60050.8)& EAVN  & 43 & 256 & $\sim110$ & $0.92\times0.47, -14.3$ &  177.7 & 1.3 & $199\pm20$ \\
2023/04/19 (60053.8) & EAVN & 43 & 256 & $\sim250$ & $0.91 \times 0.48, -20.7$ & 294.2 & 1.2  & $333\pm33$\\
%2023/04/24%\footnotemark[1]  & EAVN  & 43  & 256 & $\sim140$  & ...  & ... & ... & $118\pm59$ \\
2023/04/24 (60058.8)& EAVN  & 43  & 256 & $\sim140$  & ...  & ... & ... & $118\pm59$ \\
\hline
2023/09/30 (60217.5)& VLA & 10 & 4000 & $\sim 42$ & $160 \times 152$, 69.7 & 121.6 & 0.08 & $173\pm17$\\ 
\hline
\end{tabular*} \\
{
%Column\,(1): project code.
Column\,(1): observation date.
Column\,(2): stations participating for different projects: the East Asian VLBI Network (EAVN) includes VLBI Exploration of Radio Astrometry (VERA)--Mizusawa (VM), VERA--Iriki (VR), VERA--Ogasawara (VO), Yamaguchi (YM) and Hitachi (HT) in Japan; Tianma (T6) and Nanshan (UR) telescopes in China; as well as the Korean VLBI Network (KVN)--Yonsei (KY) and KVN--Ulsan (KU) stations in South Korea. The 6.7\,GHz observations mainly include VERA, HT, YM, T6, and UR. The 43\,GHz has KVN, VERA, and T6; the Very Large Array (VLA) includes 27 antennas in the U. S.;
Column\,(3): central frequency. 
Column\,(4): recording bandwidth.
Column\,(5): total on-source integration time.  
Column\,(6): size and PA of the elliptical Gaussian synthesized beam.
Columns\,(7)--(9) peak intensity, rms noise level ($1\sigma$), and total flux density of the image, respectively.
}
\end{table*}

\subsection{EAVN Observations and data reduction} \label{EAVN}
As mentioned above, we observed GRS\,1915+105 with EAVN during the flare events in April 2023. The 2023 observations were conducted at 6.7 and 43\,GHz over six epochs, with each frequency observed in three epochs (see \autoref{table:GRS1915_observation_summary}). These data were recorded in left-hand circular polarization at a rate of 1024 Mbps with 2-bit sampling, yielding a total bandwidth of 256\,MHz divided into eight intermediate frequencies (IFs). Data correlation was performed using the Daejeon hardware correlator at the Korea--Japan Correlation Center.

We calibrated the data following standard procedures using the Astronomical Image Processing System {\tt AIPS}~\citep{Greisen2003}. For phase calibration, we first corrected for the parallactic angle, ionospheric fluctuations, and instrumental delays. Next, global fringe fitting was performed on both the target and calibrator sources. The priori amplitude calibration was performed using the antenna gain curves and system temperatures, with opacity corrections. A scaling factor of 1.3 was then applied to account for amplitude losses caused by issues with the Daejeon correlator \citep{Lee_2015JKAS...48..229L}. Bandpass calibration was derived from scans of bright fringe-finder sources. After completing the calibration, we imported the data of the calibrators J1925+1227 (6.7\,GHz) and J1925+2106 (43\,GHz) into the Caltech {\tt DIFMAP} package \citep{Shepherd_Difmap_1997ASPC..125...77S} for iterative amplitude and phase self-calibration. Using the resulting source models, we executed the {\tt AIPS} task {\tt CALIB} on J1925+1227 (6.7\,GHz) and J1925+2106 (43\,GHz) to derive time-dependent amplitude gain corrections for each antenna. Subsequently, these calibration solutions were applied to GRS\,1915+105. Finally, we performed self-calibration and imaging for the target using the {\tt DIFMAP}. All the epochs except the 43\,GHz observations on April 24, 2023 obtained good fringe quality and final images.

\subsection{Archival VLA Data and reduction} \label{VLA_Data}
We also analyzed archival VLA data observed in September 2023 to cover another flare event and study the evolution of the jet's PA. The September observations were performed in two bands (i.e., C and X bands), we analyzed the X-band data for high angular-resolution images, which were recorded at a central frequency of 10\,GHz with a bandwidth of about 4\,GHz (\autoref{table:GRS1915_observation_summary}). Data calibration was performed using the Common Astronomy Software Applications {\tt CASA}~\citep{McMullin_CASA} package provided by NRAO, along with the VLA calibration pipeline. 
Given the wide bandwidth of these VLA observations, we further split the data by selecting different spectral windows. This allowed us to obtain multi-frequency images at 8, 9, 10, 11, and 12\,GHz from the September observations.

\section{Results} \label{Method:Results}

\subsection{Large misalignment between continuous jet and discrete ejecta}  \label{Method:radio_morphology_and_jet_PA}
\begin{figure*}[htbp!]
\centering
    \includegraphics[width=\linewidth]{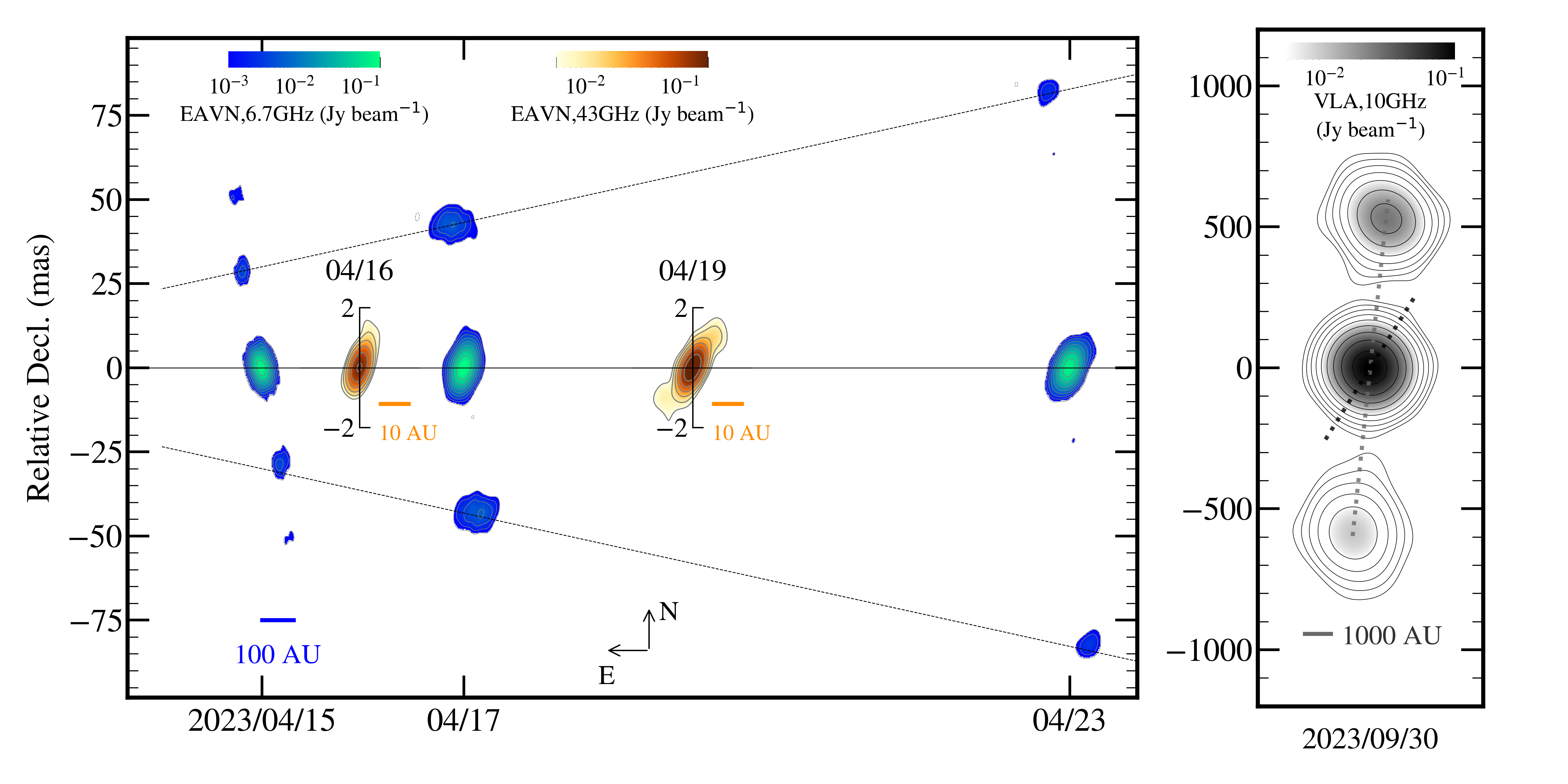}
\caption{Large misalignment of position angles between continuous jets and discrete ejecta in GRS\,1915+105. Radio images are obtained by the East Asian VLBI Network (EAVN) in the left and the Karl G. Jansky Very Large Array (VLA) in the right, with the image parameters summarized in \autoref{table:GRS1915_modelfit_EAVN_data}. At the distance of GRS\,1915+105 ($\sim9.4$\,kpc), 1\,mas corresponds to 9.4\,AU here. The two-sided blobs at both 6.7\,GHz of EAVN and 10\,GHz of VLA are recognized as discrete ejecta. The extended structure at 43\,GHz of EAVN and the center of VLA are continuous jets. The FWHM of synthesized beams are around $6\,\text{mas}\times3\,\text{mas}$ and $0.9\,\text{mas}\times0.5\,\text{mas}$ for the EAVN at 6.7\,GHz (blue) and 43\,GHz (red), respectively. It is about $160\,\text{mas}\times150\,\text{mas}$ for the VLA at 10\,GHz (grey). Contours in the three frequencies start at four times the image rms noise levels of 0.8, 5.0, and 0.3\,mJy\,beam$^{-1}$, respectively, and increase by factors of two. The PA directions of main two-sided ejecta (S1 and N1) from EAVN images at 6.7\,GHz: PA $\sim 192^\circ$, $\sim 186^\circ$, and $\sim 186^\circ$ on April 15, 2023, April 17, 2023, and April 23, 2023, respectively. The dashed lines are linear fitting to radial distances of the S1 and N1 components as a function of time (see \autoref{table:GRS1915_modelfit_EAVN_data}). The apparent speeds are $6.50 \pm 0.18$\,mas\,d$^{-1}$ for S1 and $6.53 \pm 0.19$\,mas\,d$^{-1}$ for N1, respectively. Extrapolation further indicates that both components were ejected around MJD~60045. The PA of continuous jets from EAVN 43\,GHz data: PA $\sim 158^\circ$, and $\sim 146^\circ$ on April 16, 2023 and April 19, 2023, respectively; PA $\sim 174^\circ$ and $\sim 148^\circ$ for the discrete ejecta (S2 and N2) and center continuous jet are indicated by dotted lines from VLA 10\,GHz data on September 30, 2023, respectively.}
\label{fig:GRS1915_images}
\end{figure*}

 Our timely follow-up observations with EAVN clearly resolved both a central continuous jet component and two-sided discrete ejecta (see \autoref{fig:GRS1915_images}). The positions of central component derived from phase referencing observations is consistent with that inferred from proper-motion measurements \citep{Dhawan_2007ApJ...668..430D,Reid_2014,Reid_2023ApJ...959...85R}, confirming its association with the BH of the system. In \autoref{fig:GRS1915_images}, we presented the EAVN and VLA images obtained in April and September of 2023, respectively. Our observations reveal a systematic misalignment between the PAs of the central continuous jet and downstream ejecta, indicating departures from a single, fixed jet axis. 

 We used the {\tt MODELFIT} task to model the source structure with several circular/elliptical Gaussian functions in {\tt DIFMAP}. The fitted parameters include flux density, radial distance, size, and PA of the component.
The uncertainties of these parameters were estimated following the method outlined by \citet{lee_2008AJ....136..159L}, which is primarily based on the signal-to-noise ratio (S/N). Based on the model-fitted results of the core and ejecta components, we further determined the PA of the jet ejecta through a linear fit, as shown in \autoref{fig:GRS1915_images} and in \autoref{table:GRS1915_modelfit_EAVN_data}. The three-epoch 6.7\,GHz data yield a mean jet PA of $188^\circ \pm 3^\circ$. Notably, \citet{Rodr_2025ApJ...986..108R} have reported a PA of $174^\circ \pm 1^\circ$ for a similar N--S jet observed after the August flare, indicating a significant deviation from the typical PA of $\sim 147^\circ$ for the SE--NW jet. 

\begin{table*}[]
\caption{Model-fitted Parameters of Gaussian components for EAVN and VLA data} \label{table:GRS1915_modelfit_EAVN_data}%
\begin{tabular*}{\textwidth}{@{\extracolsep\fill}ccccccccccc}
%\begin{tabular}{@{}cccccccccccc@{}}
\toprule
$\nu$ & Epoch & Comp. & $r$ & $\text{PA}_{\rm comp}$  & $S_{\nu}$ & $\Theta_{\rm major}$ & Ratio & ${\rm PA}$ & ${\rm VA}$ \\ %& $\mu_{r}$ \\
(GHz) & & & (mas) & (deg) & (mJy) & (mas) & & (deg) & (deg) & \\ %& (mas\,d$^{-1}$)\\
 (1) & (2)& (3)& (4)& (5) & (6) & (7) & (8) & (9) & (10) \\
\midrule
\multirow{9}{*}{6.7} & \multirow{3}{*}{60049.8} & core & $0.0\pm0.3$ & 5 & $449\pm45$ & $4.0\pm0.2$ & 0.77 & \multirow{3}{*}{$192\pm1$}   \\
&  & S1 & $29.9\pm0.6$ & $191\pm1$ & $13\pm5$ & $3.4\pm1.2$ & 1 &  \\
&  & N1 & $29.5\pm0.8$ & $12\pm2$ & $15\pm6$ & $4.0\pm1.6$ & 1 &  \\

\cmidrule(l){2-9}
& \multirow{3}{*}{60051.8} & core & $0.0\pm0.5$ & 2 & $365\pm37$ & $4.1\pm0.2$ & 0.76 &  \multirow{3}{*}{$186\pm1$}  \\
&  & S1 & $43.1\pm1.0$ & $187\pm1$ & $22\pm7$ & $6.2\pm2.1$ & 1 &  & $89\pm1$  \\
&  & N1 & $43.3\pm1.2$ & $6\pm2$ & $20\pm8$ & $6.3\pm2.4$ & 1 &    \\

\cmidrule(l){2-9}
&  \multirow{3}{*}{60057.8}&core & $0.0\pm0.5$ & $174$ & $281\pm28$ & $4.2\pm0.2$ & 0.78 &  \multirow{3}{*}{$186\pm1$} \\
&  & S1 & $81.9\pm0.9$ & $186\pm1$ & $10\pm4$ & $3.5\pm1.7$ & 1 &  \\
&  & N1 & $81.9\pm0.9$ & $5\pm1$ & $11\pm4$ & $3.9\pm1.8$ & 1 &  \\

\hline
\multirow{5}{*}{43} & \multirow{2}{*}{60050.8}  & core & $0.0\pm0.1$ & $157$ & $182\pm22$ & $0.2\pm0.1$ & 0.33 & \multirow{2}{*}{$158\pm7$} &\multirow{5}{*}{...} \\
&  & {NW} & $1.1\pm0.2$ & $338\pm11$ & $13\pm7$ & $0.5\pm0.4$ & 1 &    \\ 
\cmidrule(l){2-9}
& \multirow{3}{*}{60053.8} & core & $0.0\pm0.1$ & $148$ & $305\pm31$ & $0.2\pm0.1$ & 0.45 &  \multirow{3}{*}{$146\pm4$}  \\
&  & SE1 & $1.3\pm0.2$ & $144\pm10$ & $22\pm11$ & $0.9\pm0.2$ & 1 &  \\
&  & NW1 & $1.2\pm0.2$ & $328\pm7$ & $27\pm11$ & $0.7\pm0.3$ & 1 &   \\
%\cmidrule(l){2-9}

\hline
\multirow{3}{*}{10} & \multirow{3}{*}{60217.5} & core  & $2.1\pm0.9$ & $148\pm3$ & $121\pm5$ & $33\pm6$ & 0.15 & {$148\pm3$} & \multirow{3}{*}{$87\pm3$}  \\
& & S2 & $592\pm1$ & $174\pm1$ & $16\pm3$ & $105\pm2$ & 1 &  \multirow{2}{*}{$174\pm1$}  \\
& & N2 & $534\pm1$ & $354\pm1$ & $37\pm5$ & $82\pm2$ & 1 &   \\

\botrule
\end{tabular*}
%\footnotetext
{
Column (1): frequency.
Column (2): observation MJD.
Column (3): component label.
Columns (4)--(10): radial distance,  position angle of the major axis for the elliptical Gaussian component or the center of circular component relative to the phase center, flux density, major-axis size, axial ratio, position angle of continuous jet or ejecta, and viewing angle.}
%\end{sidewaystable}
\end{table*}

Interestingly, the 43\,GHz observations detected an extended component, which is compact and comparable in angular extent to the quasi-continuous synchrotron jet previously resolved in GRS 1915+105 \citep{Dhawan_2000ApJ...543..373D}. Its similar morphology and size therefore indicate that it traces the innermost, continuously jet, rather than discrete ballistic ejecta. The PA is determined by the PA of major axis of fitted elliptical Gaussian component, and show a change in PA from $158^\circ \pm 7^\circ$ on April 16 2023 to $146^\circ \pm 4^\circ$ on April 19 2023. 
These results are summarized in \autoref{table:GRS1915_modelfit_EAVN_data}. VLBI at 43\,GHz not only provides the highest spatial resolution but also mitigates the scattering to approach the innermost central continuous jet component. Its PA of $\sim146$--$158 ^\circ$ (see \autoref{fig:GRS1915_images}) is broadly consistent with the historically inferred jet direction.  In contrast, the downstream ejecta detected at 6.7 GHz at nearby epochs are aligned along a markedly different direction, with a PA of $\sim188^\circ$ (see \autoref{fig:GRS1915_images}).  

Similarly, a reanalysis of VLA 10 GHz observations obtained in September 2023 \citep{Rodr_2025ApJ...986..108R} reveals a notable offset between the central and outer components on much larger spatial scales, with PAs of $\sim 148 ^\circ$ and $174^\circ$, respectively, albeit with a smaller angular difference. The results of this observation were presented in \citet{Rodr_2025ApJ...986..108R}, and our image in \autoref{fig:GRS1915_images} is consistent with theirs. We therefore adopt the results reported by these authors, including an ejecta PA of $174^\circ \pm 1^\circ$ and a jet VA of $87^\circ \pm 3^\circ$. However, the PA of the central region was not reported. We therefore determined the PAs of the central compact core at 8, 9, 10, 11, and 12\,GHz with elliptical Gaussian fitting. From these, we estimated a mean value of $148^\circ \pm 3^\circ$ (see \autoref{fig:GRS1915_images}). This is consistent with the long-standing jet PA $\sim147^\circ$ as well as the PAs $\sim150^\circ$ of central component detected in May and June of 2024 \citep{Rodr_2025ApJ...986..108R}. On the other hand, the spectral index of the central component is $0.1\pm0.1$ ($\propto\nu^{-\alpha}$) in frequency range of 8\,GHz to 12\,GHz, indicating it is dominated by an optically thick continuous jet rather than ejecta.

\subsection{Relatively Low Intrinsic Velocity and Large Angular Change of discrete Ejecta} \label{Method:Apparent_speed_and_VA}
The identification of the south and north components in three epochs of 6.7 GHz EAVN observations \autoref{fig:GRS1915_images} enables us to estimate its intrinsic velocity. Using the model-fitted results listed in \autoref{table:GRS1915_modelfit_EAVN_data}, we derived the apparent speeds of the jet components via a linear fitting. As shown by the dashed lines in \autoref{fig:GRS1915_images}, we obtained apparent speeds of $6.50 \pm 0.18$\,mas\,d$^{-1}$ and $6.53 \pm 0.19$\,mas\,d$^{-1}$ for the south and north components. These results indicate that the southern and northern ejecta move at essentially the same speed. Furthermore, extrapolation further indicates that both components are ejected around MJD 60045.

Based on the measured apparent speeds, we can determine the jet VA from the equation \citep{Fender_1999MNRAS.304..865F}:
\begin{equation} \label{eq:theta_jet_1}
    {\rm VA} = \tan^{-1} \left[1.16\times 10^{-2} \left(\frac{\mu_{\rm app} \mu_{\rm rec}}{\mu_{\rm app} - \mu_{\rm rec}}\right) \left(\frac{d}{1\text{kpc}}\right) \right],
\end{equation}
where $\mu_{\rm app}$ and $\mu_{\rm rec}$ are the apparent speeds of the approaching and receding jets, respectively, in units of mas\,d$^{-1}$, and $d$ is the source distance in kpc. Adopting a distance of 9.4\,kpc for GRS\,1915+105 \citep{Reid_2023ApJ...959...85R}, we derive the VA to be $89^\circ \pm 1^\circ$ for the N--S orientation jet observed in April 2023. \citet{Rodr_2025ApJ...986..108R} reported a VA of $87^\circ \pm 3^\circ$ for the similar N--S jet observed in September 2023. These results indicate that the N--S jets in both April and September 2023 were oriented nearly perpendicular to our line of sight. The apparent speed of $\sim6.5$\,mas\,d$^{-1}$ corresponds to an intrinsic velocity of $\sim0.35$\,c.

\autoref{table:GRS1915_modelfit_EAVN_data} summarizes the properties of the jets, including the central continuous jet and downstream ejecta, observed in April and September of 2023, including their PA, VA, and the corresponding changes relative to the pre-2023 SE--NW jets. By combining the changes in jet PA and VA we calculate the total change in the orientation of the jet, $\Delta \Phi_T$, using the first spherical law of cosines:
\begin{equation} \label{eq:Phi_jet_T}
\cos(\Delta \Phi_T) = \cos(\Delta {\rm PA}) \cdot \cos(\Delta {\rm VA}),
\end{equation}
As listed in \autoref{table:GRS1915_modelfit_EAVN_data}, the discrete ejecta in April and September 2023 exhibit total angle changes of $46^\circ \pm 7^\circ$ and $34^\circ \pm 7^\circ$, respectively.

From \autoref{Method:Results}, we conclude that the jet properties in April and September 2023 differ significantly from those observed since 1994 \citep{Mirabel_1994Natur.371...46M}. Specifically, the jets ejecta in April and September 2023 were aligned along the N--S direction, whereas the long-standing jet direction was oriented along the SE--NW axis \citep{Rodr_2025ApJ...986..108R}. In addition, our results reveal slowly moving ejecta with an intrinsic velocity of $0.35\,c$, significantly lower than the previously reported relativistic jet velocity \textgreater ~$0.9\,c$ \citep[by adopting VA of 64$^\circ$ and distance of 9.4 kpc, e.g. ][]{Fender_1999MNRAS.304..865F, 2003MNRAS.340.1353F, Miller-Jones_2007MNRAS.375.1087M}.

\section{Explanations and Discussions} \label{sec:disc}
\subsection{Scenario of warped disk} \label{Warp}

Since 2019, GRS 1915+105 has transitioned into a novel ``obscured phase", characterized by a significant drop in X-ray flux and heavy intrinsic absorption (column densities $N_\mathrm{H} > 10^{23}-10^{24}$ cm$^{-2}$), while remaining active in radio and infrared bands \citep[e.g. ][]{Balakrishnan_2021ApJ...909...41B,Motta_2021MNRAS.503..152M,Athulya_2023MNRAS.525..489A,Gandhi_2025MNRAS.537.1385G}. The leading interpretation for this behavior is a change in the accretion disk geometry, specifically the development of a ``puffed-up" outer disk or a significant warp that intercepts the line of sight \citep[e.g. ][]{Neilsen2020ApJ,Miller2025ApJ}. The high inclination of the radio jets observed in 2023 \citep[$\sim 90^\circ$ in this work and ][]{Rodr_2025ApJ...986..108R} implies the accretion disk was viewed edge-on, if the jet is perpendicular to the disk as usually assumed. This geometry also supports the scenario that the X-ray obscuration results from a warped or thickened outer accretion disk has moved into the line of sight, effectively blocking the inner X-ray emission region. 

As these warped or tilted disk propagate inward, general relativistic magnetohydrodynamic (GRMHD) simulations show that the disk can be teared into discrete, differentially precessing rings or sub-disks \citep[e.g. ][]{Nixon2013MNRAS,Liska2021MNRAS}. Such misaligned accretion disks may also account for the complex jet behavior during the obscured phase. In particular, simulations indicate intermittent jet ejections can be launched and propagate along the rotation axes of local sub-disks \citep[e.g. ][]{Liska2018MNRAS,Liska2021MNRAS,Chatterjee2025}. Magnetic reconnection in the corona of a warped disk may provide an additional channel for producing jet ejections \citep[e.g. ][]{Eikenberry2003astro,Yuan2009MNRAS,Shende2019ApJ,Lin2023}. In both cases, ejections are expected to follow  the local orientation of the warped disk structure. 
We think it is plausible that the N--S discrete ejecta observed in 2023 are associated with torn outer sub-disks that have tilted into an edge-on configuration, also causing the X-ray obscuration. As these tilted sub-disks propagate inward, jet ejections are episodically launched aligned with their distinct, highly inclined rotation axes.

\begin{figure}[]
    \centering
    \includegraphics[width=0.7\linewidth]{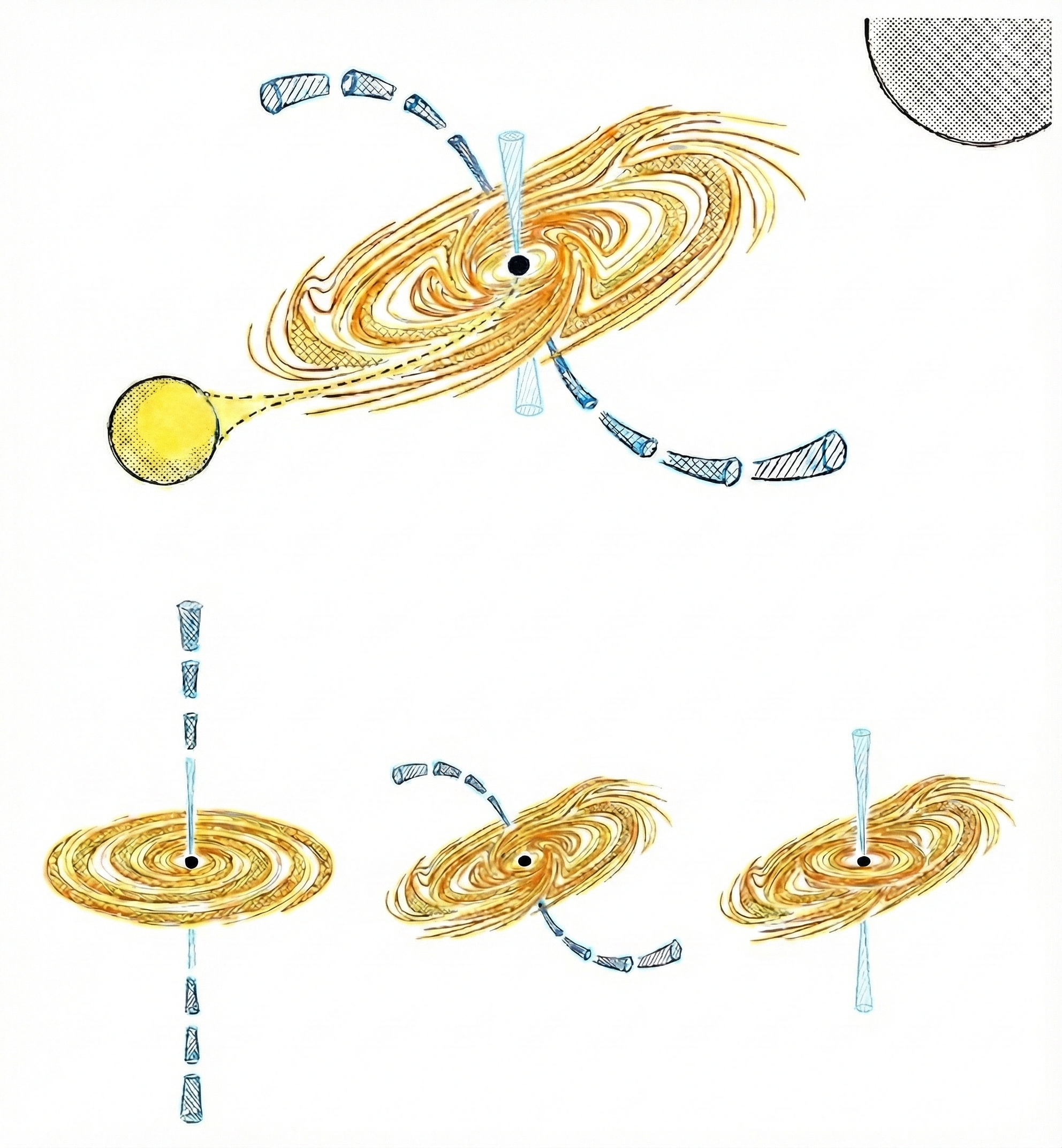}
    \includegraphics[width=0.8\linewidth]{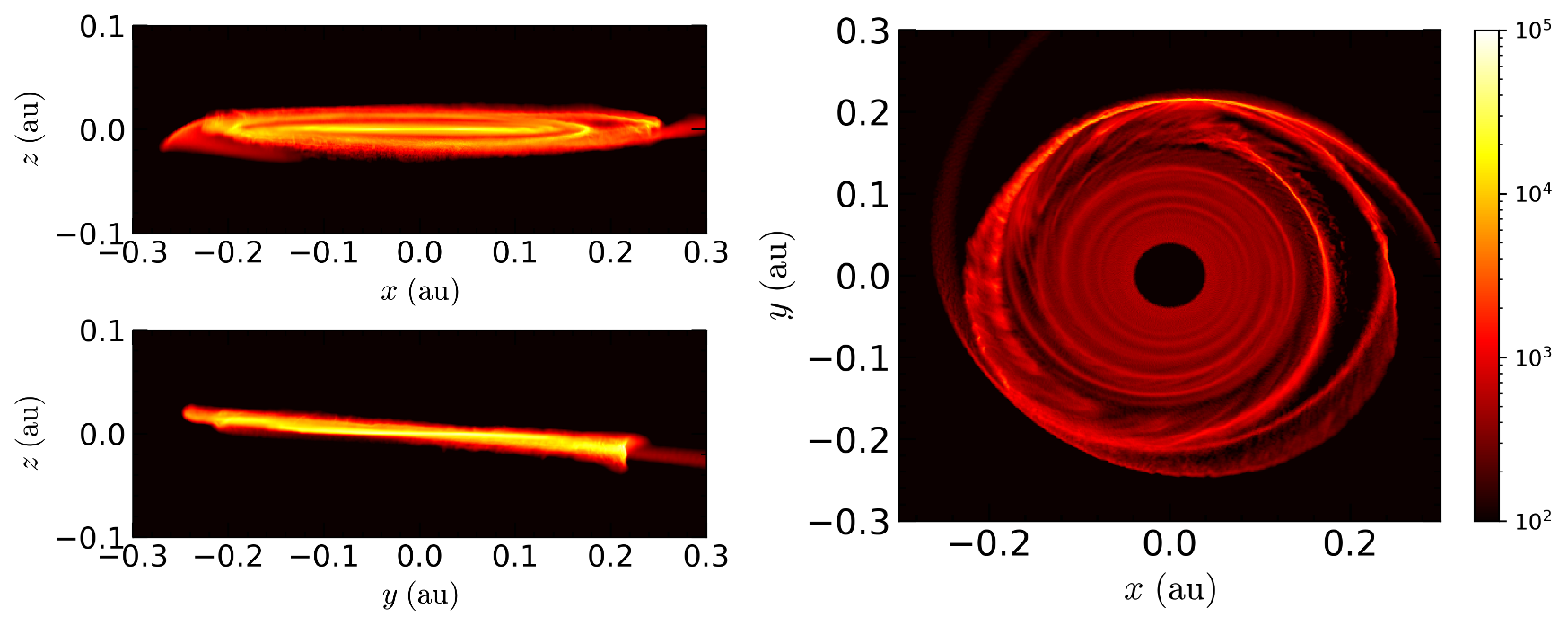}

    \caption{Top: schematic plots of warped disk and continuous jet or ejecta in GRS\,1915+105 generated with Nano Banana Pro. The long-standing jet axis is represented by the left of the second panel, the warped disk producing ejecta at a different direction is in the middle, and the inner disk aligned to BH spin producing the continuous jet is on the right. As the inner disk is aligned quickly, both the early launched discrete ejecta and the continuous jet could be seen in the same image. Bottom: the snapshots for the disk around the BH (centered at the origin) perturbed by a flyby object in a parabolic orbit. Three panels show the density contour at the $x-y$, $y-z$, and $x-y$ plane right after the object passes pericenter. }
    \label{fig:Schematic plot}
\end{figure} 

In contrast, the innermost accretion flow may decouple from the warped outer disk, and/or rapidly realign towards the black hole spin through magneto-spin alignment and/or the Bardeen-Petterson effect, reaching a dynamic equilibrium that wobbles around the spin axis  \citep{McKinney2013Sci,Liska2019MNRAS,Liska2021MNRAS,Chatterjee2025,Fender2025NatAs}. So the black hole spin provides a stable reference over very long timescales and the alignment timescale of the inner accretion flow is short, the innermost disk are expected to remain confined to a narrow range of orientations for most of the time (\autoref{fig:Schematic plot}).
The observed continuous jet is therefore likely associated with this realigned inner accretion flow. This interpretation is supported by our finding that the continuous jet, with a PA of 146--158$^\circ$, remains aligned with the historical value \citep[e.g.][]{Dhawan_2000ApJ...543..373D}, in contrast to the highly variable directions of the discrete ejecta. In this scenario, the continuous jet is preferentially tied to a time-averaged or re-established inner accretion disk that settles along the system's stable angular momentum vector (e.g., the black hole spin) after transient ejection events have occurred or become dynamically decoupled from the outer warped disk. The continuous jet is commonly observed in the hard state of BHXRBs. However, due to the lack of sufficient high angular resolution observations, it remains unclear whether their continuous jet direction is generally stable over long timescales. Our results for GRS 1915+105 suggest that, at least under some conditions, the continuous jet preferentially aligns with a stable axis, plausibly the black hole spin.

Therefore, the large misalignment between the continuous jet and discrete ejecta also offers new insight into the long-standing debate over whether continuous and transient jets in BHXRBs are produced by fundamentally different launching mechanisms \citep[e.g. ][]{Fender2004MNRAS,Fender2010MNRAS,Massi2011MmSAI}. 
Recent work by \citet{Fender2025NatAs} suggests that discrete ejecta themselves may arise from multiple mechanisms, with slower ones capable of changing their orientations and potentially being launched from the accretion disk. 
In this context, the discrete ejecta detected in this work exhibit the lowest velocity yet observed in GRS~1915+105 \citep{Miller-Jones_2007MNRAS.375.1087M} and a substantial change in orientation of $\sim46^\circ$, further supporting an accretion disk origin for at least some discrete ejecta. 
The combination of low velocity and large angular change in GRS 1915+105 directly corroborates the statistical 
trend in \citet{Fender2025NatAs}, namely that transient jets with variable directions tend to be intrinsically slower. In contrast, it is unclear from existing observations whether the aligned continuous jets have different intrinsic velocities. 

Several other BHXRBs have also exhibited an “obscured phase” characterized by strong and highly variable local absorption \citep[e.g.,][]{Koljonen2020A}. The column-density changes observed in these systems are reminiscent of X-ray changing-look active galactic nuclei (CLAGN), which undergo transitions between Compton-thin and Compton-thick states \citep[e.g.,][]{Matt2003MNRAS,Marchesi2022ApJ,Lyu2025MNRAS}. In both classes of objects, the physical nature and location of the dense medium that obscures the central X-ray emitting region, whether clumpy disk winds, warped disks or more distant torus-like structures, remain under active debate \citep[e.g.][]{Koljonen2020A,Lyu2025MNRAS}. 
Our results suggest that changes in accretion-jet geometry and X-ray obscuration may be physically coupled. An analogous case is V404 Cyg, where the black hole spin is misaligned with its binary orbital plane \citep{Khargharia2010ApJ,Miller-Jones2019Natur,Salvesen2020MNRAS}. Within our proposed framework, this spin-orbit misalignment naturally gives rise to a tilted accretion disk. The dynamic, tilted accretion flow, in turn, drives both the X-ray obscuration and the changing jet orientations observed in the system \citep{Miller-Jones2019Natur}. Furthermore, the discrete ejecta in V404 Cyg are relatively slow, exhibiting intrinsic velocities that range from $0.32\, c$ to $0.48\, c$ \citep{Miller-Jones2019Natur,Fender2025NatAs}. To date, however, there have been no reports about the large orientation changes of accretion flow or jet/outflow in X-ray CLAGNs. If a similar mechanism operates across different sources and mass scales, transient geometric distortions of the accretion flow, such as disk warping or tearing, could alter the X-ray obscuration phase and jet/outflow launching conditions. In this context, searching for radio jet structural changes, polarization-angle variations, or PA swings across X-ray unobscured and obscured phases becomes particularly compelling, as they provide a diagnostic for distinguishing between purely absorptive scenarios and those involving dynamical reconfiguration of the inner accretion flow.

\subsection{Simulation of disk warped by a perturber} \label{simulation}

The X-ray obscuration and the change in jet orientation have emerged only in recent years, despite more than three decades of continuous observations of GRS 1915+105.  The abrupt onset of these anomalies suggests the action of an external perturbation. A close flyby or the periastron passage of a tertiary object on a highly eccentric orbit could exert sufficient tidal torque to induce a strong warp in the outer disk, appearing as a transient phenomenon when compared to the system’s long observational history. Numerical studies of planetary disks demonstrate that such encounters can generate highly warped disks and even trigger disk tearing \citep[see \autoref{fig:Schematic plot} and e.g.,][]{Xiang2016MNRAS,Nealson2020MNRAS}. 

To investigate disk warping by a flyby object, we use the smoothed particle hydrodynamical (SPH) code PHANTOM \citep{Price2018} to simulate the three-dimensional gravitational interaction between a thin disk and a flyby object in a parabolic orbit (i.e. orbital eccentricity $e=1$). The effect of a tertiary object in a bound orbit ($e<1$) is qualitatively similar and will not be discussed here. Note that the companion star orbiting around the BH (with a semi-major axis of $0.47\ {\rm au}$) is neglected in our simulations for simplicity.
We setup a perturber with its mass of $3\ M_{\odot}$ and the pericenter of $0.50\ {\rm au}$. The inclination of the flyby orbit is chosen as $i=45^{\circ}$ with an argument of pericenter $0^{\circ}$. We have tested that such a flyby object will not disturb the orbit of the companion star significantly.
The disk is modelled with initially $10^{6}$ equal mass SPH particles with a total disk mass $M_{\rm d}=10^{-5}\ M_{\odot}$. These particles are initially radial distributed from the inner disk radius $r_{\rm in}=0.02\ {\rm au}$ to the outer radius $r_{\rm out}=0.3\ {\rm au}$.
The gas surface density follows a power-law distribution $\Sigma(r)\propto r^{-p}$ with $p=3/4$ for a Shakura \& Sunyaev disk. We adopt a locally isothermal equation of state and set the sound speed $c_{s}$ by the disk thickness profile as $H/r=0.015(r/1\ {\rm au})^{-0.25}$.
The Shakura \& Sunyaev viscosity \citep{SS1973}  is prescribed by $\nu=\alpha_{\rm SS}c_{s}H$ with $\alpha_{\rm SS}=0.1$.

A snapshot close to the pericenter orbital phase of our simulation is shown in the bottom panels of \autoref{fig:Schematic plot}. The three panels show the disk density viewed from the $x-z$, $y-z$ and $x-y$ plane. We can see that the outer disk is puffed up and warped due to the tidal interaction with the flyby object, with a warping amplitude as large as $20^{\circ}$. The warping propagates inward gradually by tracking the temporal evolution of the warping amplitude. Although the disk warping amplitude shown here is still less than $\sim45^{\circ}$ inferred by the misalignment between the continuous and transient jets, it would be expected that a larger misalignment could be induced by the non-linear coupling between the warped/teared disk and the transient jet \citep[e.g. ][]{Liska2018MNRAS,Liska2021MNRAS,Chatterjee2025}. The companion star, which is neglected here, could also affect the disk warping, especially its long term evolution. 

If the flyby or tertiary object is sufficiently massive and close enough to induce a substantial disk warp, it should also perturb the binary orbit of GRS 1915+105. Future high-precision astrometric observations can test this prediction by measuring deviations of proper motions as well as the binary’s orbital motion.
Current simulations are limited to a special configuration for the perturber; a systematic study of the flyby object or the tertiary object on the disk warping will be presented in detail in a separate work.

\subsection{Other explanations} \label{explanations}

While the scenario linking the obscured phase and jet reorientation of GRS 1915+105 to a warped, torn disk potentially induced by a tertiary/flyby object pericentre passage seems roughly consistent with numerical simulations, there are also several uncertainties and alternative explanations must be considered. First, the physical nature of the obscuring medium remains debated. The obscuration is highly variable and complex, occasionally becoming Compton-thick \citep[$N_\mathrm{H}>10^{24}$ cm$^{-2}$; e.g. ][]{Balakrishnan_2021ApJ...909...41B,Athulya_2023MNRAS.525..489A}. It is unclear whether the geometric warp can account for the extreme column density variations observed on short timescales. 

Some studies have instead suggested that it arises from dense disk winds enshrouding the central X-ray–emitting region \citep[e.g. ][]{Miller_2020ApJ...904...30M,Gandhi_2025MNRAS.537.1385G}. In such a scenario, clumpy and localized winds naturally explain the rapid and large-amplitude variability of the absorbing column density. Furthermore, if a sufficiently massive wind expands into the jet propagation channel, interactions between the jet and the surrounding medium could, in principle, alter the jet orientation \citep[e.g.,][]{Dubus2010MNRAS,Prabu2025arXiv}, which may account for the misalignment between the continuous jet and discrete ejecta. However, jet deflection by a dense wind typically produces non-ballistic trajectories and pronounced bending, which are inconsistent with the largely ballistic motion observed for the discrete ejecta in GRS~1915+105. Moreover, the continuous jet in GRS~1915+105 extends over spatial scales of several AU \citep[see \autoref{fig:GRS1915_images} and][]{Dhawan_2000ApJ...543..373D}, and the discrete ejecta are much further, far exceeding the binary separation. Whether disk winds can remain sufficiently dense to interact with the jet on such large scales is therefore questionable.

\section{Summary} \label{sec:conc}
In this work, we report a significant misalignment between the discrete ejecta and the continuous jet in the microquasar GRS\,1915+105, revealed by high-resolution radio interferometric observations. The continuous jet exhibits a PA broadly consistent with the historically long-standing value, while the discrete ejecta exhibited a large angular offset. This phenomenon was captured during two consecutive flare episodes in 2023, when the source was in an X-ray obscured phase. The large angular offset indicates a time-variable jet launching geometry during the obscured phase, which is different from the decades long X-ray bright stage. 
%of the transient ejection may be linked to an accretion flow that is misaligned with respect to the black hole spin. 
The abrupt onset of the jet reorientation further implies the action of an external torque. We propose a scenario in which the outer accretion disk was perturbed by a massive flyby or tertiary object. This disturbance propagates inward, warping and possibly tearing the inner disk. %at scales of several tens of gravitational radii. 
Ejecta launched from such a warped disk would naturally propagate along a different direction. The innermost accretion flow may rapidly realign towards the black hole spin via mechanisms such as magneto-spin alignment and/or the Bardeen-Petterson effect. This framework could account for the quasi-simultaneous appearance of two distinct types of jets with markedly different ejection angles in the VLBI images, as well as their simultaneous detection in VLA data. The large misalignment and low velocity of discrete ejecta also offer new insight into the long-standing debate over whether continuous and transient jets in BHXRBs are produced by fundamentally different launching mechanisms. Moreover, the transient nature and rare occurrence of such an event over the past three decades provide important clues for understanding other transient astrophysical phenomena, such as changing-look active galactic nuclei.  
%% Please use the acknowledgment and contribution environments. This will 
%% be anonomyized when the "anonymous" style option is used. 
\begin{acknowledgments}
The authors greatly appreciate the anonymous referee for very insightful suggestions in improving the manuscript. This work is supported by the National Natural Science Foundation of China (Grant Nos. 12173074, 12573100, 12373049, 12361131579, 12373070, and 12192223). 
X.Y. was supported by the China Postdoctoral Science Foundation under Grant Number 2025M773200. X.Y. also acknowledges support from the Xinjiang Tianchi Talent Program and the 2025 Outstanding Postdoctoral Grant of the Xinjiang Uygur Autonomous Region. Y.P.L. is also supported in part by the National Natural Science Foundation of Shanghai (Grant No. 23ZR1473700).
L.C. acknowledges support from the Tianshan Talent Training Program (grant No. 2023TSYCCX0099). The authors deeply appreciate the EAVN coordinator Dr. Kiyoaki Wajima for prompt actions of our ToO requests and all staff members in EAVN who rapidly initiated follow-up observations and correlated the data after the radio flares in 2023 and 2025 were reported. We thank the help of KJCC, Dr. Kazuya Hachisuka, and Dr. Shuangjing Xu for revising the delay models for phase referencing observations.

This work is made use of the East Asian VLBI Network (EAVN), which is operated under cooperative agreement by National Astronomical Observatory of Japan (NAOJ), Korea Astronomy and Space Science Institute (KASI), Shanghai Astronomical Observatory (SHAO), Xinjiang Astronomical Observatory (XAO), Yunnan Astronomical
Observatory (YNAO), National Astronomical Research Institute of Thailand (Public Organization) (NARIT), and National Geographic Information Institute (NGII), with the operational support by Ibaraki University (for the operation of Hitachi 32-m and Takahagi 32-m telescopes), Yamaguchi University (for the operation of Yamaguchi 32-m telescope), and Kagoshima University (for the operation of VERA Iriki antenna).

The KVN is a facility operated by the Korea Astronomy and Space Science Institute (KASI) and VERA is a facility operated by the National Astronomical Observatory of Japan (NAOJ) in collaboration with associated universities in Japan. T6 is operated by Shanghai Astronomical Observatory. UR is operated by Xinjiang Astronomical Observatory, and the Urumqi Nanshan Astronomy and Deep Space Exploration Observation and Research Station of Xinjiang (XJYWZ2303). HT is operated by NAOJ and Ibaraki University. YM is operated by Yamaguchi University.

The Very Long Baseline Array and the Karl G. Jansky Very Large Array are operated by the National Radio Astronomy Observatory, facilities of the U.S. National Science Foundation operated under cooperative agreement by Associated Universities, Inc.
\end{acknowledgments}

%\begin{contribution}
%%This section gives authors the space to recognize author contributions. The text inside this environment is NOT counted towards the total word quanta. At a minimum, manuscripts are expected to include this text:

%All authors contributed equally to the Terra Mater collaboration.

%% But authors are expected to provide more specific details, e.g. 
%%
%%SC was responsible for writing and submitting the manuscript.
%%WWM came up with the initial research concept and edited the manuscript.
%%OTS obtained the funding and edited the manuscript.
%%EBF provided the formal analysis and validation. He also edited the manuscript.
%%GEH Supervised the undergraduates, wrote the software and administers the project github and Zenodo repositories.
%%
%% Authors can use the Contributor Role Taxonomy (CRediT) at
%% for ideas on how write a good statement tailored to their needs.

%\end{contribution}

\facilities{EAVN, VLA}

\software{astropy \citep{Astropy2013},  AIPS, CASA, Difmap
          } \\

%%%
\appendix

\section{phase referencing calibration and results} \label{sec:pr}
\begin{figure}
\centering
\includegraphics[width=1\textwidth]{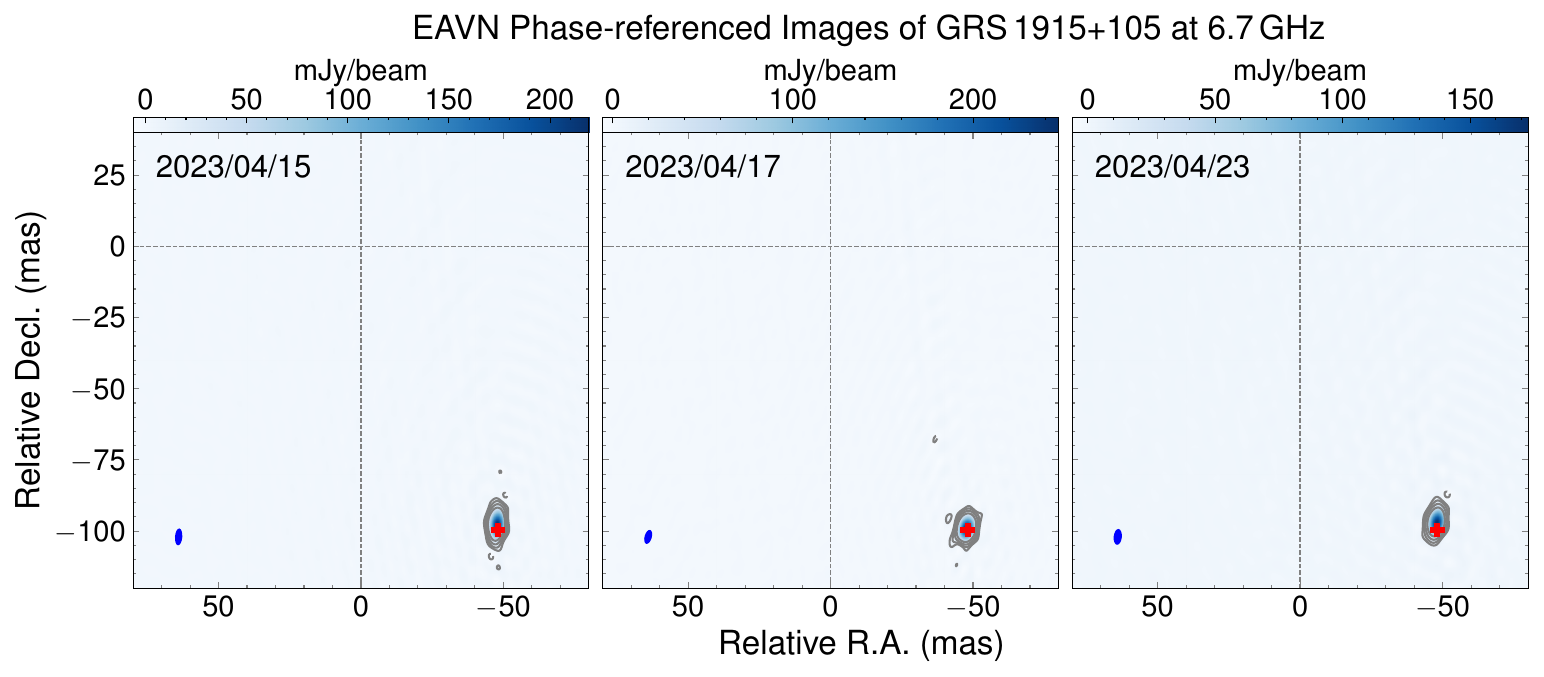}
\caption{
Phase-referenced images of GRS\,1915+105 observed with EAVN at 6.7\,GHz on April 15 (left), 17 (middle), and 23 (right), 2023. Contours start at 0.5\,mJy\,beam$^{-1}$, increasing by factors of 2. The synthesized beam is indicated in the bottom-left corner of each panel. The J2000 coordinates of the map center (0,0) are ($19^{\rm h}15^{\rm m}11\secdot5473$, $+10^{\circ}56^{'}44\arcsecdot7040$), corresponding to the source positions at an epoch within 2007--2008. The red cross marks the expected source position in 2023, ($19^{\rm h}15^{\rm m}11\secdot54409$\,mas, $+10^{\circ}56^{'}44\arcsecdot6044$\,mas), extrapolated from proper motion estimates. These results suggest the detection of a bright, compact core of GRS\,1915+105 in our phase-referenced images.}
\label{fig:GRS1915_PRed_maps}
\end{figure}

Our 6.7\,GHz observations from 2023 employed fast-switching mode between GRS\,1915+105 and the phase-referencing calibrator J1925+1227, which is separated by $2.^\circ$98 and exhibits a compact point-source structure. The switching cycle was approximately 5 minutes. Thus, we performed independent phase-referencing calibrations on these data sets for astrometric purposes. We first read the delay recalculation table into {\tt AIPS}, which included the latest station coordinates, accurate source coordinates, the most up-to-date Earth Rotation Parameters, and corrections for tropospheric and ionospheric delays~\citep{Sakai_2023PASJ...75..208S}. The remaining steps were similar to those described in \autoref{EAVN}, except that fringe fitting was performed only on the calibrator J1925+1227, using its source model obtained from its clean image. The residual phase, delay, and rate solutions were then interpolated to the consecutive scans of GRS\,1915+105. Through these steps, we successfully obtained phase-referenced images of GRS\,1915+105, in which the relative position of GRS\,1915+105 with respect to the phase-referencing center was preserved (see \autoref{fig:GRS1915_PRed_maps}). In this analysis, we excluded the YM and HT antennas due to the uncertainties in their geodetic positions, and the UR telescope due to significant phase fluctuations of scattering at long baselines. 

For our EAVN observations in 2023, we adopted the J2000 coordinates of GRS\,1915+105, ($19^{\rm h}15^{\rm m}11\secdot5473$, $+10^{\circ}56^{'}44\arcsecdot7040$), as the phase-referencing center. This corresponds to the source position at an epoch within 2007--2008 \citep{Reid_2014}. Given the source's proper motion, it is necessary to determine the accurate core positions of GRS\,1915+105 in 2023 prior to conduct further analyses.

Using an 8.4\,GHz VLBA phase-referencing image obtained in 2019 and the {\tt JMFIT} task in {\tt AIPS}, we derived the updated positions for GRS\,1915+105 on MJD~58768. Combining with the source position on January 1 2000 (MJD~51544) adopted from \citet {Dhawan_2007ApJ...668..430D}, we derived the following proper motions for GRS\,1915+105:
%\begin{align*}
$\mu_{\rm R.A.} = -3.18\pm0.73~\rm mas~yr^{-1} $, and $\mu_{\rm Decl.} = -6.15\pm0.99~\rm mas~yr^{-1} $. We note that \citep{Reid_2023ApJ...959...85R} reported variance-weighted average proper motions of $\mu_{\rm R.A.} = -3.14 \pm 0.03\rm~mas\,yr^{-1}$ and $\mu_{\rm Decl.} = -6.23 \pm 0.04\rm~mas\,yr^{-1}$, based on results from \citep{Dhawan_2007ApJ...668..430D} and \citep{Reid_2014}. More recently, \citep{Rodr_2025ApJ...986..108R} reported values of $\mu_{\rm R.A.} = -3.46 \pm 0.48\rm~mas\,yr^{-1}$ and $\mu_{\rm Decl.} = -5.94 \pm 0.48\rm~mas~yr^{-1}$. Our proper motion measurements are consistent with these previously reported results. Furthermore, the phase-referenced images of 2023 EAVN observations at 6.7\,GHz are presented in \autoref{fig:GRS1915_PRed_maps}, which reveal a single bright component. The positions of this component were derived using the {\tt MODELFIT} task in {\tt DIFMAP}. Notably, the positions agree excellently with the expected source positions extrapolated from proper motion estimates. This confirms that the bright, compact core of GRS\,1915+105 is detected during the April 2023 flare event and its association with the black hole of the system. 

%% For this sample we use BibTeX plus aasjournalv7.bst to generate the

\bibliography{GRS1915_MS}{}
\bibliographystyle{aasjournalv7}



\end{document}